\documentclass[article,oneside,a4paper]{memoir}
\usepackage[T1]{fontenc}
\usepackage[utf8]{inputenc}
\usepackage{lmodern}
\usepackage{amsmath,amsfonts,amsthm,mathrsfs,paralist,amssymb,bm,amsthm,graphicx,color,url,mathrsfs,xcolor,biblatex}

\usepackage[all]{xy}

\bibliography{sampling.bib}

\newcommand{\addpoint}[1]{#1\ ---\ }

\setsecnumdepth{subsubsection}

\setsecnumformat{\csname the#1\endcsname---}
\setsubsechook{\setsecnumformat{\csname the##1\endcsname\ ---\ }}
\setsechook{\setsecnumformat{\csname the##1\endcsname---}}

\setsecheadstyle{\centering\Large\bfseries\sffamily}
\setsubsecheadstyle{\large\bfseries\sffamily}
\setsubsubsecheadstyle{\itshape\addpoint}
\setsubsubsecindent{1em}
\setbeforesubsubsecskip{0em}
\setsubsubsechook{\setsecnumformat{{\normalfont\csname
      the##1\endcsname\ }}}
\setaftersubsubsecskip{-0em}
\setsubparaheadstyle{\itshape}
\newtheoremstyle{thm}
     {1.5ex plus .3ex minus .1ex}
     {1ex plus .3ex minus .1ex}
     {\itshape}
     {}
     {\sffamily}
     {---}
     {0em}
     {$\bullet$\hbox{\ }#1\hbox{\ }#2}

\theoremstyle{thm}

\newtheorem{definition}{Definition}
\newtheorem{theorem}{Theorem}

\newtheorem{lemma}{Lemma}
\newtheorem{proposition}{Proposition}

\newtheoremstyle{note}
     {1ex plus .3ex minus .1ex}
     {1ex plus .3ex minus .1ex}
     {}
     {}
     {\itshape}
     {.}
     {1em}
     {}
\theoremstyle{note}

\newlength{\remaining}

\newcommand{\verticale}{\ar@{--}[d]}

\def\calli#1{\expandafter\def\csname
  #1\endcsname{\mathcal{#1}}}	
\def\sets#1{\expandafter\def\csname
  bb#1\endcsname{\mathbb{#1}}}	
\def\rebar#1{\expandafter\def\csname #1bar\endcsname{\overline{\csname
      #1\endcsname}}}		
\def\gothify#1{\expandafter\def\csname
  #1#1#1\endcsname{\mathfrak{#1}}}	

\sets{R}	
\sets{N}	
\sets{Z}	
\calli{I}	
\calli{M}	
\calli{T}	
\calli{R}
\calli{U}
\calli{W}
\gothify{F}	
\gothify{G}	
\rebar{M}	

\newcommand{\FFFbar}{\overline{\FFF}}
\newcommand{\RR}{\mathscr{R}}

\newcommand{\C}{\mathscr{C}}

\newcommand{\up}[1]{\,\uparrow #1}
\newcommand{\Up}[1]{\,\Uparrow #1}

\newcommand{\pr}{\mathbb{P}}
\newcommand{\esp}{\mathbb{E}}

\newcommand{\Cstar}{\mathfrak{C}}

\newcommand{\slgb}{\mbox{$\sigma$-al}\-ge\-bra}

\newcommand{\BM}{\partial\M}

\newcommand{\tq}{\;:\;}

\def\calli#1{\expandafter\def\csname #1\endcsname{\mathcal{#1}}}
\def\sets#1{\expandafter\def\csname bb#1\endcsname{\mathbb{#1}}}
\def\rebar#1{\expandafter\def\csname #1bar\endcsname{\overline{\csname
      #1\endcsname}}}
\calli{I}
\calli{M}
\calli {T}
\sets{N}
\sets{R}
\rebar{M}
\rebar{phi}

\newcommand{\height}{\tau}
\newcommand{\cut}[1]{\kappa_{#1}}
\newcommand{\uup}{\Up}
\newcommand{\N}{\mathbb{N}}
\newcounter{mylocalcounter}
\DeclareMathOperator{\Id}{Id}

\begin{document}

\title{\huge\bfseries Uniform Generation in Trace
  Monoids\thanks{Published in the proceedings of \emph{Mathematical Foundations of
      Computer Science 2015, Milano}}}
\author{Samy Abbes\thanks{Universit\'e Paris Diderot/PPS CNRS UMR 7126
  and IRISA/INRIA CNRS UMR 6074 \texttt{samy.abbes@univ-paris-diderot.fr}}\and Jean Mairesse\thanks{UPMC/LIP6 CNRS UMR 7606 \texttt{jean.mairesse@lip6.fr}}}
\date{2014}
\maketitle

 \begin{abstract}
   We consider the problem of random uniform generation of traces (the
   elements of a free partially commutative monoid) in light of the
   uniform measure on the boundary at infinity of the associated
   monoid. We obtain a product decomposition of the uniform measure at
   infinity if the trace monoid has several irreducible components---a
   case where other notions such as Parry measures, are not
   defined. Random generation algorithms are then examined.
\par\vspace{.5em}\hspace{-3.2em}\textbf{Keywords:} trace monoid, uniform
  generation, M\"obius polynomial
 \end{abstract}

\section{Introduction}
\label{sec:introduction}

Uniform generation of finite-size combinatorial objects consists in
the design of a randomized algorithm that takes an integer $k$ as
input, and returns an object of size~$k$, such that each object of
size $k$ has equal probability to be produced. This problem has been
considered for many classes of objects from computer science or
discrete mathematics: words, trees, graphs are examples. Several
general approaches exist: recursive methods~\cite{FZV94}, the Markov
chain Monte-Carlo method with coupling from the past~\cite{jerrum13},
or the Boltzmann sampler~\cite{duchon04}.  Other recent approaches
share a common guideline, namely first considering a notion of
\emph{uniform measure on infinite objects} in order to gain,
afterwards, information on the uniform distributions on finite
objects. The theory of random planar graphs is an example of
application of this idea.  In this paper, we investigate the uniform
generation of traces (elements of a trace monoid) and we base our
approach on the notion of {\em uniform measure on infinite traces}.

Given an independence pair $(A,I)$, where $I$ is an irreflexive and
symmetric relation on the finite alphabet~$A$, the associated trace
monoid $\M=\M(A,I)$ contains all congruence classes of the free monoid
$A^*$, modulo equivalences of the form $ab=ba$ for all $(a,b)\in I$,
see \cite{cartier69,diekert90}. Elements of $\M$ are called
\emph{traces}. Trace monoids are ubiquitous in Combinatorics,
see~\cite{viennot86}. They are also one of the most basic models of
concurrency under a partial order semantics~\cite{diekert95}. Uniform
generation of traces is thus a fundamental question with possible
applications in probabilistic model checking of concurrent
systems. Since our concern is with partial order semantics, it differs
from the sequential approach which targets uniform generation of
linear executions in models of concurrency~\cite{bodini12}.

Consider a trace monoid~$\M$, and, for each integer $k\geq0$, the
finite set $\M_k=\{x\in\M\tq|x|=k\}$. Let $\nu_{\M_k}$ be the uniform
distribution over $\M_k$. A crucial observation is that the
probability measures $(\nu_{\M_k})_{k\in \N}$ are \emph{not}
consistent. Consequently, the uniform measures $\nu_{\M_k}$ cannot be
reached by a recursive sampling of the form $x_1\cdot\ldots\cdot
x_k\in\M_k$\,, with the $x_i$'s being sampled independently and
according to some common distribution over~$A$.

To overcome the difficulty, several steps are necessary.  First, we
consider the \emph{uniform measure at infinity} for~$\M$, a notion
introduced in \cite{abbesmair14} for irreducible trace monoids, and
extended here to the general case.  Second, we prove a realization
result for the uniform measure at infinity by means of a Markov chain
on a combinatorial sub-shift. Last, we apply the results to the
uniform sampling of finite traces.  None of the three steps is
straightforward. Besides standard uniform sampling, it turns out that
evaluating the uniform average cost or reward associated with traces
can be done in an efficient way.

An original feature of our approach is to define the measure at
infinity for general trace monoids and not only for irreducible
ones. We show that the uniform measure at infinity of a
\emph{reducible} trace monoid decomposes as a product of measures on
irreducible components---contrasting with uniform distribution at
finite horizon. In general, the uniform measure at infinity charges
the \emph{infinite} traces of the ``largest'' components of the
monoid, and charges the \emph{finite} traces of the ``smallest''
components.

Another, different but related, notion of `uniform measure' exists:
the Parry measure which is a uniform measure on bi-infinite sequences
of an \emph{irreducible} sofic sub-shift~\cite{parry64,kitchens98}.
The construction can be applied to trace monoids, defining a `uniform
measure' on bi-infinite traces, but only for irreducible trace
monoids.  Here we focus on single sided infinite traces instead of
bi-infinite ones, and this approach allows to relax the irreducibility
assumption, and to construct a uniform measure at infinity for a
general trace monoid. In case the trace monoid is irreducible, we
provide a precise comparison between the Parry measure, restricted to
single sided infinite traces, and our uniform measure at infinity. The
latter turns out to be a non-stationary version of the former. Another
important point is that our approach reveals the combinatorial
structure hidden in the uniform measure at infinity (and in the Parry
measure).


The outline of the paper is the following. We first focus in a warm-up
section (\S~\ref{sec:warm-up:-uniform}) on the case of two commuting
alphabets.  Relaxing the commutativity assumption, we arrive to trace
monoids in~\S~\ref{sec:unif-gener-trac}. The purpose of
\S~\ref{sec:unif-gener-finite} is twofold: first, to compare the
uniform measure with the Parry measure; and second, to examine
applications to the uniform sampling of finite traces.

\section{Warm-up: uniform measure for commuting alphabets}
\label{sec:warm-up:-uniform}

Let $A$ and $B$ be two alphabets and let $\M$ be the product monoid
$\M=A^*\times B^*$.  The size of $u=(x,y)$ in $\M$ is
$|u|=|x|+|y|$. Let $\partial A^*=A^{\N}$ be the set of infinite
$A$-words, let $\overline{A^*}=A^* \cup A^{\N}$, and similarly
for~$\partial B^*$ and $\overline{B^*}$\,.  Define:
\begin{align*}
\BM &= \bigl\{(\xi,\zeta) \in \overline{A^*}\times \overline{B^*} \; :
\; |\xi|+|\zeta|=\infty \bigr\}\,,&
 \Mbar&=\M\cup\BM\,. 
\end{align*}
Clearly one has $\BM = (\overline{A^*}\times\overline{B^*} ) -
(A^*\times B^*)$ and $\Mbar=\overline{A^*}\times\overline{B^*}$.
Both $\overline{A^*}$ and $\overline{B^*}$ are equipped with the
natural prefix orderings, and $\Mbar$ is equipped with the product
ordering, denoted by~$\leq$. For $u\in\M$, we put: 
\begin{align*}
\uup
u&=\{v\in\Mbar\tq u\leq v\}\,, &
\up u&=\{\xi\in\BM\tq u\leq \xi\}\,.
\end{align*}

Let $p_0=1/|A|$ and $q_0=1/|B|$\,. Without loss of generality, we assume
that $|A|\geq|B|$, hence $p_0\leq q_0$\,.  

\begin{lemma}
  \label{lem:1}
  For each real number $p\in (0,p_0]$, there exists a unique
  probability measure $\nu_p$ on~$\overline{A^*}$ such that\/
  $\nu_p(\uup x)=p^{|x|}$ holds for all $x\in A^*$\,. We have: 
\begin{align*}
  \forall p \in (0,p_0)\qquad \nu_p (A^*) &=1\,, & \nu_{p_0} (\partial
  A^*) &=1 \,.
\end{align*}
\end{lemma}

The probability measures $\nu_p$ in Lemma~\ref{lem:1} are called
\emph{sub-uniform} measures of parameter~$p$
over~$\overline{A^*}$. The measure $\nu_{p_0}$ is the classical
uniform measure on $\partial A^*$ which satisfies $\nu_{p_0}(\up
x)=p_0^{|x|}$ for all $x\in A^*$.

For each integer $k\geq0$, let $\nu_{\M_k}$ denote the uniform
distribution on $\M_k=\{(x,y)\in\M\tq |(x,y)|=k\}$\,. Since
$|A|\geq|B|$, an element $(x,y)\in\M_k$ sampled according to
$\nu_{\M_k}$ is more likely to satisfy $|x|\geq|y|$ than the opposite.
In the limit, it is natural to expect that infinite elements on the
$B$ side are not charged at all, except if $|A|=|B|$. This is made
precise in the following result.

\begin{theorem}
  \label{thr:4}
  Let $\nu_A$ and $\nu_B$ be the sub-uniform measures of parameter
  $p_0=1/|A|$ over $\overline{A^*}$ and $\overline{B^*}$ respectively.
  The sequence $(\nu_{\M_k})_{k\geq0}$ converges weakly to the product
  measure \mbox{$\nu=\nu_A\otimes\nu_B$}.

  We have: $\nu\bigl(\up(x,y)\bigr)=p_0^{|x|+|y|}$ for all\/
  $(x,y)\in\M$; and\/ $\nu(\partial A^*\times B^*)=1$ if\/ $|A|>|B|$,
  whereas $\nu(\partial A^*\times \partial B^*)=1$ if\/ $|A|=|B|$.
%
%
\end{theorem}

We say that the measure $\nu$ described in Th.~\ref{thr:4} is the
\emph{uniform} measure on~$\BM$. We have the following ``realization''
result for~$\nu$.

\begin{theorem}
\label{thr:5}
Let $(a_n)_{n\in \N}$ be a sequence of i.i.d.\ and uniform random
variables (r.v.) over~$A$.  Let $b_0$ be a r.v.\ over
$B\cup\{1_{B^*}\}$\,, where $1_{B^*}$ is the identity element
of~$B^*$, and with the following law:
\begin{align*}
\forall b\in B\quad  \pr(b_0 = b)&= p_0=1/|A|\,,&
\pr(b_0=1_{B^*})&=1- p_0/q_0=1- |B|/|A|\,.
\end{align*}

Consider $(b_n)_{n\in\bbN}$ sampled independently in\/
$B\cup\{1_{B^*}\}$\,, each $b_n$ with the same law as~$b_0$\,, but
only until it reaches\/~$1_{B^*}$\,, after which $b_n$ is constant
equal to\/~$1_{B^*}$\,.
%
Finally, set $u_k\in\M$ for all integers $k\geq0$ by:
\begin{align*}
  x_k&=a_0\cdot\ldots\cdot a_{k-1}\in A^*\,,&y_k&=b_0\cdot\ldots\cdot
  b_{k-1}\in B^*\,,&u_k&=(x_k,y_k)\in\M\,.
\end{align*}

Then $(u_k)_{k \in \N}$
converges in law towards~$\nu$\,. Furthermore, the random variable
$\bigvee_{k\geq0}u_k\in\BM$ is distributed according to~$\nu$.
\end{theorem}


Observe that $1_{B^*}$ will eventually appear in the sequence
$(b_n)_{n \in \N}$ with probability~$1$ if and only if $p_0<q_0$. In
this case, $(y_n)_{n\in \N}$ is eventually equal to a constant element
of $B^*$ with probability~$1$. This is consistent with
Theorem~\ref{thr:4}.  Observe also that $(a_n,b_n)_{n \in \N}$ forms a
product Markov chain on \mbox{$A\times(B\cup\{1_{B^*}\})$}.

Both results stated in Ths.~\ref{thr:4} and~\ref{thr:5} are particular
cases of corresponding results for trace monoids, as we will see next.

\section{Uniform and sub-uniform measures for trace monoids}
\label{sec:unif-gener-trac}

\paragraph{\bfseries Basics on trace monoids.}

Let $A$ be a finite alphabet equipped with an irreflexive and
symmetric relation $I\subseteq A\times A$, called an
\emph{independence relation}. The pair $(A,I)$ is called an
\emph{independence pair}. Let $\I$ be the congruence relation on the
free monoid $A^*$ generated by the collection of pairs $(ab,ba)$ for
$(a,b)$ ranging over~$I$.  The \emph{trace monoid\/} $\M=\M(A,I)$ is
defined as the quotient monoid $\M=A^*/\I$, see
\cite{cartier69,viennot86,diekert90}. The elements of $\M$ are called
\emph{traces}. The identity element in the monoid is called the
\emph{empty trace}, denoted~``$1_{\M}$'', and the concatenation is
denoted with the dot~``$\cdot$''\,.

The length of a trace $u$ is well defined as the length of any of its
representative words and is denoted by $|u|$. The left divisibility
relation on $\M$ is a partial order, denoted by ``$\leq$'' and defined
by: $u\leq v\iff \exists w\quad v=u\cdot w$\,.

%

\begin{figure}[t]
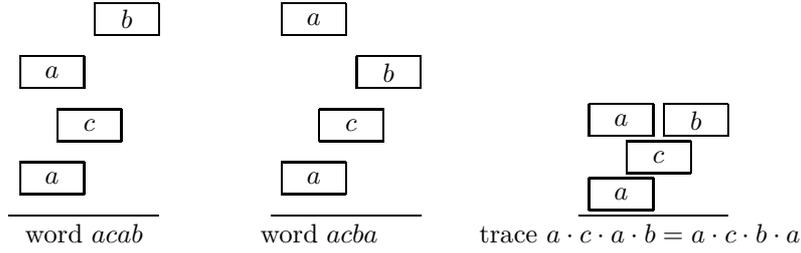

  \centering
  \begin{tabular}{ccc}
\xy
<.1em,0em>:
(0,6)="G",
"G"+(12,6)*{a},
"G";"G"+(24,0)**@{-};"G"+(24,12)**@{-};"G"+(0,12)**@{-};"G"**@{-},
(14,26)="G",
"G"+(12,6)*{c},
"G";"G"+(24,0)**@{-};"G"+(24,12)**@{-};"G"+(0,12)**@{-};"G"**@{-},
(0,46)="G",
"G"+(12,6)*{a},
"G";"G"+(24,0)**@{-};"G"+(24,12)**@{-};"G"+(0,12)**@{-};"G"**@{-},
(28,66)="G",
"G"+(12,6)*{b},
"G";"G"+(24,0)**@{-};"G"+(24,12)**@{-};"G"+(0,12)**@{-};"G"**@{-},
(-4,-2);(52,-2)**@{-}
\endxy
& \quad\qquad
\xy
<.1em,0em>:
(0,6)="G",
"G"+(12,6)*{a},
"G";"G"+(24,0)**@{-};"G"+(24,12)**@{-};"G"+(0,12)**@{-};"G"**@{-},
(14,26)="G",
"G"+(12,6)*{c},
"G";"G"+(24,0)**@{-};"G"+(24,12)**@{-};"G"+(0,12)**@{-};"G"**@{-},
(0,66)="G",
"G"+(12,6)*{a},
"G";"G"+(24,0)**@{-};"G"+(24,12)**@{-};"G"+(0,12)**@{-};"G"**@{-},
(28,46)="G",
"G"+(12,6)*{b},
"G";"G"+(24,0)**@{-};"G"+(24,12)**@{-};"G"+(0,12)**@{-};"G"**@{-},
(-4,-2);(52,-2)**@{-}
\endxy&\qquad \xy
<.1em,0em>:
0="G",
"G"+(12,6)*{a},
"G";"G"+(24,0)**@{-};"G"+(24,12)**@{-};"G"+(0,12)**@{-};"G"**@{-},
(14,14)="G",
"G"+(12,6)*{c},
"G";"G"+(24,0)**@{-};"G"+(24,12)**@{-};"G"+(0,12)**@{-};"G"**@{-},
(0,28)="G",
"G"+(12,6)*{a},
"G";"G"+(24,0)**@{-};"G"+(24,12)**@{-};"G"+(0,12)**@{-};"G"**@{-},
(28,28)="G",
"G"+(12,6)*{b},
"G";"G"+(24,0)**@{-};"G"+(24,12)**@{-};"G"+(0,12)**@{-};"G"**@{-},
(-4,-2);(52,-2)**@{-}
\endxy\\
word $acab$&\quad word $acba$&\quad trace $a\cdot c\cdot a\cdot b=a\cdot c\cdot b\cdot a$
\end{tabular}
\caption{Two congruent words and the resulting heap (trace)}
  \label{fig:heapm1}
\end{figure}

An intuitive representation of traces is given by Viennot's \emph{heap
  of pieces} interpretation of a trace monoid~\cite{viennot86}.  We
illustrate in Fig.~\ref{fig:heapm1} the heap of pieces interpretation
for the monoid $\M(A,I)$ with $A=\{a,b,c\}$ and
$I=\{(a,b),(b,a)\}$.

The length of traces corresponds to the number of pieces in a
heap. The relation $u\leq v$ corresponds to $u$ being seen at bottom
as a sub-heap of heap~$v$.

The product monoid $A^*\times B^*$
from~\S~\ref{sec:warm-up:-uniform} is isomorphic to the trace monoid
$\M(\Sigma,I)$, where $\Sigma=A\cup B$ with $A$~and $B$ being
considered as disjoint, and $I= (A\times B)\cup (B\times A)$\,.


\paragraph{\bfseries Cliques and height of traces.}

Recall that a \emph{clique} of a graph is a complete subgraph (by
convention, the empty graph is a clique). We may view $(A,I)$ as a
graph.  Given a clique $c$ of $(A,I)$, the product
$a_1\cdot\ldots\cdot a_j\in\M$ is independent of the enumeration
$(a_1,\ldots,a_j)$ of the vertices composing~$c$.  We say that
$a_1\cdot\ldots\cdot a_j$ is a \emph{clique} of $\M$. Let $\C$ denote
the set of cliques, including the empty clique~$1_\M$\,.  As heaps of
pieces, cliques correspond to flat heaps, or horizontal layers.


Traces are known to admit a canonical normal form, defined as
follows~\cite{cartier69}. Say that two non-empty cliques $c,c'$ are
\emph{Cartier-Foata admissible}, denoted by $c\to c'$, whenever they
satisfy: $\forall a\in c'\quad\exists b\in c\quad (b,a)\notin I$. 
For every non empty trace $u\in\M$, there exists a unique integer
$n>0$ and a unique sequence $(c_1,\ldots,c_n)$ of non-empty cliques
such that:
\begin{inparaenum}[(1)]
  \item $u=c_1\cdot\ldots\cdot c_n$; and
  \item $c_i\to c_{i+1}$ holds for all $i\in\{1,\ldots,n-1\}$.
\end{inparaenum}
The integer $n$ is called the \emph{height} of~$u$, denoted by
$n=\height(u)$. By convention, we put $\height(1_{\M})=0$. The
sequence $(c_1,\ldots,c_n)$ is called the \emph{Cartier-Foata normal
  form} or \emph{decomposition} of~$u$. In the heap interpretation,
the normal form corresponds to the sequence of horizontal layers that
compose a heap~$u$, and the height $\height(u)$ corresponds to the
number of horizontal layers.

A useful device is the notion of \emph{topping} of traces, defined as
follows: for each integer $n\geq0$, the \emph{$n$-topping} is the
mapping $\cut n:\M\to\M$ defined by $\cut n(u)=c_1\cdot\ldots\cdot
c_n$\,, where $c_1\to\ldots\to c_p$ is the Cartier-Foata decomposition
of~$u$, and where $c_i=1_{\M}$ if $i>p$.

\paragraph{\bfseries Boundary. Elementary cylinders.}

Let $\Cstar=\C\setminus\{1_\M\}$ denote the set of non-empty
cliques. Traces of $\M$ are in bijection with finite paths of the
automaton $(\Cstar,\to)$, where all states are both initial and
final. Denote by~$\BM$ the set of infinite paths in the automaton
$(\Cstar,\to)$. We call $\BM$ the \emph{boundary at infinity}, or
simply the \emph{boundary}, of monoid~$\M$, and we put
$\Mbar=\M\cup\BM$. Elements of~$\BM$ are called \emph{infinite
  traces}, and, by contrast, elements of $\M$ might be called
\emph{finite traces}. 

By construction, an infinite trace is given as an infinite
sequence $\xi=(c_1,c_2,\ldots)$ of non-empty cliques such that $c_i\to
c_{i+1}$ holds for all integers $i\geq1$. Note that the topping
operations extend naturally to $\cut n:\Mbar\to\M$, defined by $\cut
n(\xi)=c_1\cdot\ldots\cdot c_n$\,, for $\xi=(c_1,c_2,\ldots)$. 

We wish to extend the partial order relation $\leq$ from $\M$
to~$\Mbar$. For this, we first recall the following
result~\cite[Cor.~4.2]{abbesmair14}: for $u,v\in\M$, if
$n=\height(u)$, then \mbox{$u\leq v\iff u\leq \cut n(v)$}.
Henceforth, we put $\zeta\leq\xi\iff \forall n\geq0\quad \cut
n(\zeta)\leq\cut n(\xi)$ for $\zeta, \xi \in\Mbar$, consistently with
the previous definition in case $\zeta,\xi\in\M$. This order is
coarser than the prefix ordering on sequences of cliques.

For each $u\in\M$, we define two kinds of \emph{elementary cylinders of
  base~$u$}:
\begin{align}
\label{eq:2}
  \up u&=\{\xi\in\BM\tq u\leq\xi\}\subseteq\BM\,,&
\uup u&=\{v\in\Mbar\tq u\leq v\}\subseteq\Mbar\,.
\end{align}

The set $\M$ being countable, it is equipped with the discrete
topology. The set $\Mbar$ is a compactification of~$\M$, when equipped
with the topology generated by the opens of $\M$ and all
cylinders~$\uup u$\,, for $u$ ranging over~$\M$. This makes $\Mbar$ a
metrisable compact space~\cite{abbes06}. The set $\BM$ is a closed subset
of~$\Mbar$. The induced topology on $\BM$ is generated by the family
of cylinders~$\up u$, for $u$ ranging over~$\M$. Finally, both spaces
are equipped with their respective Borel \slgb s, $\FFF$~on $\M$ and
$\FFFbar{}$ on~$\Mbar$; the \slgb\ on each space is generated by the
corresponding family of cylinders.


\paragraph{\bfseries Möbius polynomial. Principal root. Sub-uniform measures.}

We recall~\cite{cartier69,viennot86} the definitions of the
\emph{Möbius polynomial} $\mu_\M(X)$ and of the \emph{growth series\/}
$G(X)$ associated to~$\M$:
\begin{align}
  \label{eq:1}
  \mu_\M(X)&=\sum_{c\in\C}(-1)^{|c|}X^{|c|}\,,&
G(X)&=\sum_{u\in\M}X^{|u|}=\sum_{n\geq0}\lambda_\M(n)X^n\,,
\end{align}
where $\lambda_\M(n)=\#\{x\in\M\tq|x|=n\}$. It is known that $G(X)$ is
rational, inverse of the Möbius polynomial: 
\[
G(X)=1/\mu_\M(X)\:.
\]
It
is also known~\cite{krob03,csikvari13} that $\mu_\M(X)$ has
a unique root of smallest modulus, say~$p_0$\,, which lies in the real
interval $(0,1)$ if $|A|>1$ (the case $|A|=1$ is trivial). The root
$p_0$ will be called the \emph{principal root} of~$\mu_\M$\,, or
simply of~$\M$.

The following result, to be compared with Lemma~\ref{lem:1}, adapts
the so-called Patterson-Sullivan construction from geometric group
theory. The compactness of $\Mbar$ is an essential ingredient of the
proof for the case $p=p_0$\,, based on classical results from
Functional Analysis.

\begin{theorem}
  \label{thr:6}
  For each $p\in (0,p_0]$, where $p_0$ is the principal root
  of\/~$\M$, there exists a unique probability measure $\nu_p$ on
  $(\Mbar,\FFFbar{})$ such that $\nu_p(\uup x)=p^{|x|}$ holds for all
  $x\in\M$. On the one hand, if $p<p_0$\,, then $\nu_p$ is
  concentrated on~$\M$, and is given by:
  \begin{gather}
    \label{eq:24}
    \forall x\in\M\qquad \nu_p\bigl(\{x\}\bigr)=p^{|x|}/G(p)\,.
  \end{gather}

  On the other hand, $\nu_{p_0}$ is concentrated on the boundary, hence
  $\nu_{p_0}(\BM)=1$. In this case, $\nu_{p_0}(\up x)=p_0^{|x|}$ holds
  for all $x\in\M$.
\end{theorem}



\begin{definition}
  \label{def:4}
  The measures $\nu_p$ on\/ $\Mbar$ described in
  Th.~{\normalfont\ref{thr:6}} are called\/ \emph{sub-uniform measures
    of parameter~$p$}. The measure $\nu_{p_0}$ is called the\/
  \emph{uniform measure} on\/~$\BM$.
\end{definition}
 
The following result relates the uniform measure on the boundary with
the sequence $\nu_{\M_k}$ of uniform distributions over the sets
$\M_k=\{x\in\M\tq|x|=k\}$. 

\begin{theorem}
  \label{thr:7}
  Let $\M$ be a trace monoid, of principal root~$p_0$\,. The sequence
  of uniform distributions $(\nu_{\M_k})_{k\geq0}$ converges weakly
  toward the uniform measure $\nu_{p_0}$ on\/~$\BM$.
\end{theorem}

Anticipating on  Th.~\ref{thr:8} below, Theorem~\ref{thr:7}
above has the following concrete consequence. Fix an integer $j\geq1$,
and draw traces of length $k$ uniformly at random, with $k$
arbitrarily large. Then the $j$ first cliques of the trace obtained
approximately behave as if they were a Markov chain
$(C_1,\ldots,C_j)$; and the larger~$k$, the better the
approximation. Conversely, how this can be exploited for random
generation purposes, is the topic of
Sect.~\ref{sec:unif-gener-finite}.


\paragraph{\bfseries Irreducibility and irreducible components.}

Generators of a trace monoid only have partial
commutativity properties. The following definition isolates
the parts of the alphabet that enjoy full commutativity.

\begin{definition}
  \label{def:2}
  Let\/ $(A,I)$ be an independence pair. The associated\/
  \emph{dependence pair} is $(A,D)$ where $D=(A\times A)\setminus
  I$. The connected components of the graph $(A,D)$ are called the
  \emph{irreducible components} of $\M=\M(A,I)$. To each of these
  irreducible component $A'$ is associated the independence relation
  $I'=I\cap(A'\times A')$. The corresponding trace monoids
  $\M'=\M(A',I')$ are called the \emph{irreducible components} of the
  trace monoid $\M$. If\/ $(A,D)$ is connected, then $\M$ is
  said to be \emph{irreducible}.
\end{definition}


Direct products of trace monoids are trace monoids themselves. More
precisely, the following result holds.

\begin{proposition}
  \label{prop:1}
  Let $\M=\M(A,I)$ be a trace monoid. Then $\M$ is the direct product
  of its irreducible components. As a measurable space and as a
  topological space, $\Mbar$ is the product of the~$\overline{\M'}$,
  where $\M'$ ranges over the irreducible components of\/~$\M$. The
  Möbius polynomial $\mu_\M(X)$ is the product of the Möbius
  polynomials~$\mu_{\M'}(X)$, for $\M'$ ranging over the irreducible
  components of\/~$\M$.
\end{proposition}

The sets $\M_k=\{x\in\M\tq |x|=k\}$ do not enjoy a product
decomposition with respect to irreducible components of~$\M$, hence
neither do the uniform distributions $\nu_{\M_k}$ over~$\M_k$\,. By
contrast, sub-uniform measures have a product decomposition, as stated
below.

\begin{proposition}
\label{thr:1}
Let $\M$ be a trace monoid, of principal root~$p_0$\,, and let $\nu_p$
be a sub-uniform measure on $\Mbar$ with $p\leq p_0$\,.  Then
$\nu_{p}$ is the product of measures $\nu'$ on each of
the~$\overline{\M'}$, for $\M'$ ranging over the irreducible
components of~$\M$. The measures $\nu'$ are all sub-uniform measures
on $\overline{\M'}$ of the same parameter~$p$\,.
\end{proposition}

It follows from Prop.~\ref{prop:1} that the principal root of a trace
monoid $\M$ is the smallest among the principal roots of its
irreducible components.  As a consequence of Prop.~\ref{thr:1}, the
uniform measure is a product of \emph{sub-uniform} measures $\nu'$
over the irreducible components $\M'$ of~$\M$.  By Th.~\ref{thr:6},
each $\nu'$ is either concentrated on $\M'$ if the principal root $p'$
of $\M'$ satisfies $p'>p_0$\,, or concentrated on\/ $\BM'$ if\/
$p'=p_0$\,. Note that at least one of these sub-uniform measures is
actually \emph{uniform} on the irreducible component.

\paragraph{\bfseries Realization of uniform and sub-uniform measures.}

The characterization of the uniform measure by $\nu(\up x)=p_0^{|x|}$
(see Th.~\ref{thr:6}) does not provide an obvious recursive procedure
for an algorithmic approximation of $\nu$-generated samples on~$\BM$.
Since the uniform measure $\nu$ is, according to Prop.~\ref{thr:1}, a
product of sub-uniform measures, it is enough to focus on the
algorithmic sampling of \emph{sub-uniform measures on irreducible
  trace monoids}.

Hence, let $\M$ be an irreducible trace monoid, of principal
root~$p_0$\,, and let $\Mbar$ be equipped with a sub-uniform
measure~$\nu_p$ with $p\leq p_0$\,. Recall from Th.~\ref{thr:6} that
$\nu_p$ is either concentrated on $\M$ or on $\BM$ according to
whether $p<p_0$ or $p=p_0$\,. 

Elements of $\M$ are given as finite paths in the graph
$(\Cstar,\to)$, whereas elements of $\BM$ are given as infinite paths
in $(\Cstar,\to)$. In order to have a unified presentation of both
spaces, we use the following technical trick: instead of considering
the graph of \emph{non empty} cliques $(\Cstar,\to)$, we use the graph
of \emph{all} cliques $(\C,\to)$, including the empty clique. We keep
the same definition of the Cartier-Foata relation~`$\to$' (see
above). Note that $c\to 1_{\M}$ then holds for every clique $c\in\C$,
whereas $1_{\M}\to c$ holds if and only if $c=1_{\M}$. Hence $1_{\M}$
is an absorbing state in $(\C,\to)$. Any path in $(\Cstar,\to)$,
either finite or infinite, now corresponds to a unique infinite path
in $(\C,\to)$. If the original path $(c_k)_{1\leq k\leq N}$ is finite,
the corresponding infinite path $(c'_k)_{k\geq1}$ in $(\C,\to)$ is
defined by $c'_k=c_k$ for $1\leq k\leq N$ and $c'_k=1_{\M}$ for all
$k>N$.

For each trace $\xi\in\Mbar$, either finite or infinite, let
$(C_k)_{k\geq1}$ be the infinite sequence of cliques corresponding to
the infinite path in $(\C,\to)$ associated with~$\xi$. The sequence
$(C_k)_{k\geq1}$ is a random sequence of cliques; its characterization
under a sub-uniform measure $\nu_p$ is the topic of next result.

\begin{theorem}
  \label{thr:8}
  Let $\M$ be an irreducible trace monoid of principal
  root~$p_0$\,. Then, with respect to the sub-uniform measure $\nu_p$
  on\/~$\Mbar$, with\/ \mbox{$0<p\leq p_0$}\,, the sequence of random
  cliques $(C_k)_{k\geq1}$ is a Markov chain with state space~$\C$.

  Let $g,h:\C\to\RR$ be the functions defined by:
  \begin{align}
\label{eq:18}
h(c)&=\sum_{c'\in\C\tq c'\geq c}(-1)^{|c'|-|c|}p^{|c'|}\,,&
g(c)&=h(c)/p^{|c|}\,.
  \end{align}

  Then $\bigl(h(c)\bigr)_{c\in\C}$ is a probability vector over~$\C$,
  which is the distribution of the initial clique~$C_1$\,. This vector
  is positive on\/~$\Cstar$, and $h(1_{\M})>0$ if and only if $p<p_0$\,.
  The transition matrix of the chain, say
  $P=(P_{c,c'})_{(c,c')\in\C\times\C}$\,, is:
\begin{gather}
\label{eq:19}
  P_{c,c'}=
  \begin{cases}
    0,&\text{if $c\to c'$ does not hold,}\\
h(c')/g(c),&\text{if $c\to c'$ holds,}
  \end{cases}
\end{gather}
with the line $(P_{1_{\M},c'})_{c'\in\C}$ corresponding to the empty clique
undefined if $p=p_0$\,.

Conversely, if $p\leq p_0$\,, and if $(C_k)_{k\geq1}$ is a Markov
chain on $\C$ if $p<p_0$\,, respectively on $\Cstar$ if $p=p_0$\,,
with initial distribution $h$ defined in\/~\eqref{eq:18} and with
transition matrix $P$ defined in\/~\eqref{eq:19}, and if\/
$Y_k=C_1\cdot\ldots\cdot C_k$\,, then\/ $(Y_k)_{k\geq1}$ converges weakly
towards the sub-uniform measure~$\nu_p$\,. Furthermore,
the law of the random trace $C_1\cdot
C_2\cdot\ldots=\bigvee_{k\geq1}Y_k\in\Mbar$ is the probability measure
$\nu_p$ on\/~$\Mbar$.
\end{theorem}

Theorem~\ref{thr:8} for $p=p_0$ already appears in~\cite{abbesmair14}. 
Note: the function $h:\C\to\RR$ defined in~(\ref{eq:18}) is the
\emph{Möbius transform} in the sense of Rota~\cite{rota64,aigner07} of
the function $f:c\in\C\mapsto p^{|c|}$\,;
see~\cite{abbesmair14} for more emphasis on this point of
view.

As expected, we recover the results of \S~\ref{sec:warm-up:-uniform}
in the case of two commuting alphabets $A$ and~$B$ with
$|A|>|B|$.  Indeed, by Prop.~\ref{thr:1} and Th.~\ref{thr:8}, the
cliques $(C_k)_{k\geq1}$ form a product of two Markov chains: one
on~$A$ (\emph{non empty cliques} of~$A^*$) and the other one on
$B\cup\{1_{B^*}\}$ (\emph{cliques} of~$B^*$, including the empty one).

\section{Uniform Generation of Finite Traces}
\label{sec:unif-gener-finite}

We have introduced in Def.~\ref{def:4} a notion of uniform measure on
the boundary of a trace monoid.  This measure is characterized by its
values on cylinders in Th.~\ref{thr:6}, as the weak limit of uniform
distributions in Prop.~\ref{thr:1}, and through the associated
Cartier-Foata probabilistic process in Th.~\ref{thr:8}.

Because of the existence of the Cartier-Foata normal form of traces,
the combinatorics of a trace monoid is entirely contained in the
Cartier-Foata automaton, either $(\C,\to)$ or $(\Cstar,\to)$. 
Looking at the Cartier-Foata automaton, say $(\Cstar,\to)$ on non
empty-cliques, as generating a sub-shift of finite type, it is
interesting to investigate the associated notion of uniform measure `à
la Parry'~\cite{parry64,kitchens98,LiMa}, and to compare it with the
uniform measure on the boundary previously introduced. This comparison
between the two notions of uniform measures will enlighten the
forthcoming discussion on uniform generation of finite traces.

\paragraph{\bfseries Uniform measure on the boundary versus Parry
  measure.}
\label{sec:unif-meas-parry}

The Parry measure associated with an \emph{irreducible} sub-shift of
finite type is formally defined as the unique measure of maximal
entropy on \emph{bi-infinite} admissible sequences of states of the
sub-shift. It corresponds intuitively to the ``uniform measure'' on
such bi-infinite paths (see, e.g.,~\cite{LiMa}).

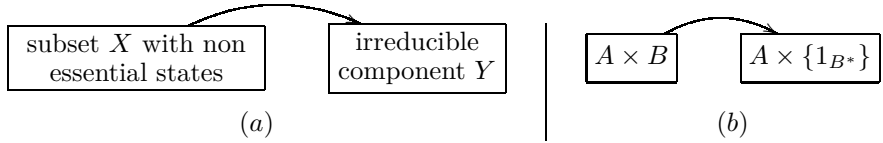
\begin{figure}[t]
  \begin{gather*}
\begin{array}{cc|cc}
\xymatrix{*+[F]{\txt<9em>{subset $X$ with non essential states}}\ar@/^2em/[r]&
*+[F]{\txt<6em>{irreducible component $Y$}}}
&\quad&\quad&
  \xymatrix{*+[F]{\txt{\strut$A\times B$}}\ar@/^1.5em/[r]&
*+[F]{\txt{\strut$A\times\{1_{B^*}\}$}}}
\\[1.3em]
(a)&&&(b)
\end{array}
\end{gather*}
\caption{\textsl{$(a)$\ Illustration of a reducible system\quad $(b)$\
    Cartier-Foata automaton on non-empty cliques of $A^*\times B^*$}
  generating the uniform measure on $\partial(A^*\times B^*)$ if
  $|A|>|B|$}
  \label{fig:pokapoapa}
\end{figure}

The Parry measure is only defined for \emph{irreducible} sub-shifts
for good reasons. Indeed, if a sub-shift has, say, two parts $X$
and~$Y$, with $Y$ an irreducible component and such that going from
$X$ to $Y$ is possible but not the other way around as in
Fig.~\ref{fig:pokapoapa}--$(a)$, then one cannot define a ``uniform
measure'' on bi-infinite paths (it should put mass on paths spending
an infinite amount of time both in $X$ and $Y$ and be stationary,
which is impossible).  On the other hand, considering a uniform
measure on one-sided infinite sequences on such a compound system
makes perfect sense. This is the case, for instance, of the
Cartier-Foata sub-shift associated to the reducible trace monoids
$A^*\times B^*$ with $|A|>|B|$ studied
in~\S~\ref{sec:warm-up:-uniform}: see Fig.~\ref{fig:pokapoapa}--$(b)$.

For a general trace monoid~$\M$, the associated sub-shift
$(\Cstar,\to)$ is irreducible if and only if the monoid $\M$ is
irreducible in the sense of Def.~\ref{def:2} (a well-known result: see
for instance~\cite[Lemma~3.2]{krob03}). Therefore the comparison
between the uniform measure on the boundary, and the Parry measure,
only makes sense in this case.

Hence, let $\M$ be an irreducible trace monoid, of principal
root~$p_0$\,.  In order to take into account the length of cliques in
the construction of the Parry measure, we consider the weighted
incidence matrix $B=(B_{x,y})_{(x,y)\in\Cstar\times\Cstar}$ defined by
$B_{x,y}=p_0^{|y|}$ if $x\to y$ holds and by $B_{x,y}=0$ if $x\to y$
does not hold.

\begin{lemma}
  \label{lem:2}
  The non-negative matrix $B$ has spectral radius\/~$1$. The vector
  $g=(g(c))_{c\in\Cstar}$ defined by $g(c)=\sum_{c'\in\Cstar\tq c\to
    c'}h(c')$ for $c\in\Cstar$\,, where $h$ has been defined
  in\/~\eqref{eq:18}, is $B$-invariant on the right: $Bg=g$.
\end{lemma}


Define the matrix 
$C=(C_{c,c'})_{(c,c')\in\Cstar\times\Cstar}$ by:
\begin{gather}
  \label{eq:20}
\forall c,c'\in\Cstar\qquad C_{c,c'}=B_{c,c'}\,g(c')/{g(c)}\,.
\end{gather}

Since $g$ is right invariant for~$B$, it follows that $C$ is stochastic. 
Classically, the Parry measure on bi-infinite paths in $(\Cstar,\to)$ is the stationary Markovian measure 
of transition matrix $C$. 

\begin{proposition}
\label{prop:2}
The matrix\/ $C$ defined in\/~\eqref{eq:20} 
coincides with the transition matrix $P$ defined in
Theorem\/{\normalfont~\ref{thr:8}} for $p=p_0$\,, and restricted to\/
$\Cstar\times\Cstar$.
\end{proposition}

Proposition~\ref{prop:2} asserts that the Markov chain associated with
the Parry measure has the same transition matrix as the probabilistic
process on non-empty cliques generated by the uniform measure on the
boundary. But the Parry measure is stationary whereas the uniform
measure $\nu$ is not. Indeed, the initial distribution of the Markov
measure $\nu$ is $h:\Cstar\to\RR$, which does \emph{not} coincide with
the stationary measure of the chain (except in the trivial case of a
free monoid).

To summarize: the notion of uniform measure on the boundary is adapted
to one-sided infinite heaps, independently of the irreducibility of
the trace monoid under consideration. If the monoid is irreducible,
there is a notion of uniform measure on two-sided infinite heaps,
which correspond to a weighted Parry measure. Considering the
projection of this Parry measure to one-sided infinite heaps, and
conditionally on a given initial clique, it coincides with the uniform
measure at infinity since they share the same transition matrix. But
the two measures globally differ since their initial measures differ.

\paragraph{\bfseries Uniform generation of finite traces, 0.}
\label{sec:bfser-unif-gener-1}

The Parry measure is a standard tool for a special type of uniform
generation. Indeed, it provides an algorithmic way of sampling finite
sequences of a fixed length~$k$, and uniformly \emph{if the first and
  the last letters of the sequence are given}. In our framework,
besides the fact that the Parry measure is only defined for an
irreducible trace monoid, it also misses the primary target of
generating finite traces of a given length~$k$ among \emph{all traces}
of length~$k$.

\paragraph{\bfseries Uniform generation of finite traces, 1.}
\label{sec:bfser-unif-gener}

Consider the problem, given a fixed integer $k>1$ and a trace
monoid $\M=\M(A,I)$, of designing a randomized algorithm which produces a
trace $x\in\M$ of length~$k$, uniformly among traces of length~$k$. 
%
%
Sub-uniform measures on the trace monoid $\M$ allow to adapt to our
framework the technique of Boltzmann samplers~\cite{duchon04} for
solving this problem.

Consider a parameter $p\in(0,p_0)$\,, where $p_0$ is the principal
root of~$\M$, and let $\xi\in\M$ be sampled according to the
sub-uniform measure~$\nu_p$\,. We have indeed $|\xi|<\infty$ with
probability~$1$ by Th.~\ref{thr:6}. Furthermore, Prop.~\ref{thr:1}
shows that $\nu_p$ decomposes as a product of sub-uniform measures of
the same parameter~$p$, over the irreducible components of~$\M$. For
each component, sampling is done through usual Markov chain generation
techniques since both the initial measure and the transition matrix of
the chain of cliques are explicitly known by Th.~\ref{thr:8}.

The algorithm is then the following: if $|\xi|=k$,
then keep~$\xi$; otherwise, reject $\xi$ and sample another
trace. This eventually produces a random trace of length~$k$,
uniformly distributed in~$\M_k$\,; since $\nu_p$ is a weighted sum of
all~$\nu_{\M_k}$\,, as shown by the expression~(\ref{eq:24}).

As usual, the optimal parameter~$p$, for which the rejection probability
is the lowest, is such that: $\esp_{\nu_p}|\xi|=k$, where
$\esp_{\nu_p}(\cdot)$ denotes the expectation with respect
to~$\nu_p$\,. Ordinary computations show that $\esp_{\nu_p}|\xi|$ is
related to the derivative of the growth function by
$\esp_{\nu_p}|\xi|=pG'(p)/G(p)=-p\mu_\M'(p)/\mu_\M(p)$\,; providing an
explicit equation 
\[
k\mu_\M(p)+p\mu_\M'(p)=0\:,
\]
to be numerically
solved in~$p$.

Unfortunately, the rejection probability approaches $1$ exponentially
fast as $k$ increases, making the algorithm less and less
efficient. A standard way to overcome this difficulty would be to
consider \emph{approximate sampling}~\cite{duchon04}, consisting in
sampling traces of length approximately~$k$.

\paragraph{\bfseries Uniform generation of finite traces, 2: evaluating an average cost.}

Uniform generation is often done in order to evaluate the
expected value of a cost function. For this purpose, a more direct
approach in our framework is based on an exact integration formula
given in Th.~\ref{thr:9} below.

Let $\phi:\M_k\to\RR$ be a cost function, and consider the problem of
evaluating the expectation $\esp_{\nu_{\M_k}}(\phi)$\,, for a fixed
integer~$k$.  For each integer $k\geq0$, let:
\begin{align*}
  \M_k&=\{x\in\M\tq|x|=k\}\,,&\lambda_\M(k)&=\#\M_k\,,&
\M_{(k)}&=\{x\in\M\tq\height(x)=k\}\,.
\end{align*}

To each function $\phi:\M_k\to\RR$ defined on traces of
\emph{length}~$k$, we associate a function $\phibar:\M_{(k)}\to\RR$
defined on traces of \emph{height}~$k$, as follows:
\begin{gather}
\label{eq:15}
\forall x\in\M_{(k)}\qquad \phibar(x)=\sum_{y\in\M_k\tq y\leq
  x}\phi(y)\,.
\end{gather}

\begin{theorem}
\label{thr:9}
Let\/ $\phibar:\M_{(k)}\to\RR$ be defined as in\/~\eqref{eq:15}. Then
the following equality holds between the expectation with respect to
the uniform distribution $\nu_{\M_k}$ on $\M_k$ on the one hand, and
the expectation with respect to the uniform measure~$\nu$ on $\BM$ on
the other hand (whether $\M$ is irreducible or not):
  \begin{gather}
\label{eq:16}
    \esp_{\nu_{\M_k}}\phi=
\bigl(p_0^k\cdot\lambda_\M(k)\bigr)^{-1}\cdot 
  \esp_\nu\phibar(C_1\cdot\ldots \cdot C_k)\,.
  \end{gather}
\end{theorem}

The generation of $(C_k)_{k\geq1}$ enables us to evaluate
$\esp_\nu\phibar(C_1\cdot\ldots \cdot C_k)$ for any integer~$k$,
provided the function $\phibar$ can be efficiently computed. In turn,
this directly depends on the numbers $\theta_k(x)=\#\{y\in\M_k\tq
y\leq x\}$ of terms in the sum~(\ref{eq:15}) defining
$\phibar(x)$. The numbers $\theta_k(x)$ might be arbitrary large; for
instance $\theta_k\bigl((a\cdot b)^k\bigr)=k+1$ for $(a,b)\in
I$. However we have the following result. 

\begin{lemma}\label{le:le}
Assume that  $\M$ is irreducible. Then, there exists $C>0$ such that:
\[
\esp_\nu\theta_k(C_1\cdot\ldots\cdot C_k) \leq C \:.
\]
\end{lemma}

To see this, apply~(\ref{eq:16}) to the
constant function $\phi=1$ on~$\M_k$\,, whose associated function is
$\phibar=\theta_k$ on~$\M_{(k)}$\,, to obtain:
\begin{gather}
\label{eq:17}
\esp_\nu\theta_k(C_1\cdot\ldots\cdot C_k)=p_0^k\cdot\lambda_\M(k)\,.
\end{gather}

The terms $\lambda_\M(k)$, coefficients of the growth series
$G(X)=1/\mu_\M(X)$\,, are asymptotically equivalent to $Cp_0^{-k}$
for some constants $C>0$ if $\M$ is
irreducible~\cite{krob03}. The result in Lemma \ref{le:le} follows. 

\smallskip

Applying usual
techniques~\cite{bertoni94} to specifically retrieve all traces $y\leq
x$ of length $k=\height(x)$ is feasible in time $O(k)$ in average and
allows to compute $\phibar(x)$, and consequently to estimate the
expectation $\esp_\nu\phibar(C_1\cdot\ldots\cdot C_k)$ \textit{via}
Markov chain sampling and a Monte-Carlo algorithm.

By~(\ref{eq:17}), applying the same estimation technique to the
function $\phi=1$ yields an estimate for the normalization factor
$p_0^k\cdot\lambda_\M(k)$\,. In passing, this also yields a
Monte-Carlo estimate for the number~$\lambda_\M(k)$.  All together, we
are thus able to estimate with an arbitrary precision both terms in
the right hand member of~(\ref{eq:16}), hence yielding an accurate
estimation of
$\esp_{\nu_{\M_k}}\phi$\,.

To summarize: generating the first $k$ layers of 
traces under the uniform measure on the boundary allows to compute the
\emph{expectation} of an arbitrary computable cost function
$\phi:\M_k\to\bbR$\,, if $\M$ is irreducible. The same applies at the
cost of a greater complexity if $\M$ is not irreducible.

\printbibliography

\clearpage
\renewcommand{\appendixpagename}{Appendix: proofs}
\appendixpage*\appendix

\noindent\textbf{Proof of Theorem~\ref{thr:6}.}
We first recall that two traces $x,y\in\M$ having a common upper
bound have a least upper bound $x\vee y$, in which case $x$ and $y$
are said to be \emph{compatible}. Therefore:
\begin{gather*}
  \uup x\,\cap\uup y=\begin{cases}
\emptyset,&\text{if $x,y$ are not compatible,}\\
\uup (x\vee y),&\text{if $x,y$ are compatible.}
  \end{cases}
\end{gather*}

It follows that the collection of elementary cylinders $\uup x$, to
which is added the empty set, is stable by finite intersections. They
form thus a $\pi$-system. It follows that a measure $\nu$ on
$(\Mbar,\FFFbar{})$ is entirely determined by its values on all
elementary cylinders~$\uup x$, for $x\in\M$. The uniqueness stated in
the theorem follows. 

We also recall that, since the growth series $G(X)$ has positive
coefficients on the one hand, and since the formal equality
$G(X)=1/\mu_\M(X)$ holds on the other hand, the radius of convergence
of the power series $G(z)$ is exactly~$p_0$\,.

Let $\nu_p$ be the probability measure on $\Mbar$ defined, for
$p\in[0,p_0)$, by:
\begin{gather*}
  \nu_p=\frac1{G(p)}\sum_{y\in\M}p^{|y|}\delta_y\,,
\end{gather*}
where $\delta_y$ denotes the Dirac measure concentrated
on~$\{y\}$. The measure $\nu_p$ is well defined for $0\leq p<p_0$
since $G(p)<+\infty$, as recalled above.

We prove that $\nu_p(\uup x)=p^{|x|}$ for all $x\in\M$. Observe that
the mapping $y\in\M\mapsto x\cdot y$ is a bijection onto $\uup
x\cap\M$. Since $\nu_p$ is concentrated on~$\M$, we compute for any
$x\in\M$:
\begin{align*}
  \nu_p(\uup x)&=\nu_p(\uup x\cap \M)\\
&=\frac1{G(p)}\sum_{y\in\M\tq y\geq x}p^{|y|}\\
&=\frac1{G(p)}\sum_{y\in\M}p^{|x\cdot y|}\\
&=p^{|x|}\frac1{G(p)}\sum_{y\in\M}p^{|y|}\\
&=p^{|x|}
\end{align*}
This shows that $\nu_p(\uup x)=p^{|x|}$ holds for all $x\in\M$.

For $p=p_0$, we adapt the construction of the so-called
Patterson-Sullivan measure. Consider any weak limit,
say~$\nu_{p_0}$\,, of $(\nu_p)_p$ by letting $p\to p_0$\,. Such a
limit exists since $\Mbar$ is compact and therefore any sequence of
probabilities on $\Mbar$ has a weakly convergent
subsequence. In~$\Mbar$, any cylinder $\uup x$ is both open and
closed, therefore its topological boundary is empty, and therefore has
null $\nu_{p_0}$-measure. By the Porte-manteau theorem (see for
instance Billingsley's \emph{Convergence of probability measures}), we
have thus:
\begin{align}
\label{eq:4}
  \nu_{p_0}(\uup x)=\lim_{p\to p_0}\nu_p(\uup x)=\lim_{p\to p_0}p^{|x|}=p_0^{|x|}\,.
\end{align}

For the same reasons, we have for every $x\in\M$:
\begin{align*}
  \nu_{p_0}\bigl(\{x\}\bigr)&=\lim_{p\to p_0}\nu_p\bigl(\{x\}\bigr)\\
&=\lim_{p\to p_0}\frac{p^{|x|}}{G(p)}\\
&=0\,,
\end{align*}
since $G(p_0)=+\infty$. Since $\M$ is countable, it follows that
$\nu_{p_0}(\M)=0$, and thus $\nu_{p_0}$ is concentrated on the
boundary. Finally, since $\up x=\uup x\cap\BM$, and using that
$\nu_{p_0}(\M)=0$, we obtain:
\begin{align*}
  \nu_{p_0}(\up x)=\nu_{p_0}(\uup x\cap \BM)=\nu_{p_0}(\uup x)=p_0^{|x|}\,.
\end{align*}
This completes the proof of Theorem~\ref{thr:6}.

\par\bigskip\noindent\textbf{Proof of Theorem~\ref{thr:7}.}\quad
Since $\Mbar$ is compact, to show that $(\nu_{\M_k})_k$ converges
toward~$\nu_{p_0}$\,, it suffices to show that $\nu_{p_0}$ is the weak
limit of any weakly convergent subsequence of~$(\nu_{\M_k})_k$\,. Let
$\nu$ be the weak limit of a weakly convergent subsequence
$(\nu_{\M_{k_j}})_j$\,.  Using the estimate
$\#\M_k\sim_{k\to\infty}Ck^Np_0^{-k}$
(see~\cite{krob03})\,, we have for all $x\in\M$ and for all
$j$ large enough:
\begin{align*}
  \nu_{\M_{k_j}}(\uup x)&=\frac1{\#\M_{k_j}}\#\{y\in\M\tq |y|=k_j\wedge
  y\geq x\}\\
  &=\frac1{\#\M_{k_j}}\#\{y\in\M\tq |y|=k_j-|x|\}\\
  &\sim_{j\to\infty}\frac1{C(k_j)^Np_0^{-k_j}}C(k_j-|x|)^Np_0^{|x|-k_j}\\
  &\to_{j\to\infty}p_0^{|x|}\,.
\end{align*}

Using the Porte-manteau theorem as in the proof of Th.~\ref{thr:6}, we
have thus:
\begin{gather*}
  \nu(\uup x)=\lim_{j\to\infty}\nu_{\M_{k_j}}(\uup x)=p_0^{|x|}\,,
\end{gather*}
and therefore $\nu=\nu_{p_0}$\,, using as above that measures are
entirely determined by their values on elementary cylinders.  This
shows that $(\nu_{\M_k})_k$ converges toward~$\nu_{p_0}$\,. The proof
of Theorem~\ref{thr:7} is complete.

\par\bigskip\noindent\textbf{Proof of Proposition~\ref{prop:1}.}\quad
For simplicity, assume that $\M$ has exactly two irreducible
components $\M_1$ and~$\M_2$\,, and let $A_1$ and $A_2$ be the
corresponding irreducible components of~$A$.

Let $i_1:\M_1\to\M$ and $i_2:\M_2\to\M$ be the natural injections, and
let $f:\M_1\times\M_2\to\M$ be defined by $f(x_1,x_2)=i_1(x_1)\cdot
i_2(x_2)=i_2(x_2)\cdot i_1(x_1)$. Then it is clear that $f$ is an
isomorphism. The natural morphism $\pi_1:\M\to\M_1$ for instance, is
entirely determined by:
\begin{gather*}
  \forall x\in A_1\cup A_2\qquad\pi_1(x)=
  \begin{cases}
    0,&\text{if $x\in A_2$\,,}\\
x,&\text{if $x\in A_1$\,.}
  \end{cases}
\end{gather*}

We extend $f$ to $\Mbar_1\times\Mbar_2\to\Mbar$ as follows. Let us
first identify $\M_1$ and $\M_2$ with sub-monoids of~$\M$,
through the natural injections $i_1$ and~$i_2$\,. The two following
properties are obvious:
\begin{enumerate}[(a)]
\item\label{item:4} $\forall x_1\in\M_1\quad\forall x_2\in\M_2\quad x_1\cdot
  x_2=x_2\cdot x_1$
\item\label{item:5} $\forall u,v,w\in\M\quad u\geq v\implies w\cdot
  u\geq w\cdot v$
\end{enumerate}

Now let $(x,y)\in\Mbar_1\times\Mbar_2$\,. Let $(c_n)_{n\geq1}$ and
$(d_n)_{n\geq1}$ be the Cartier-Foata sequence of cliques of $\M_1$
and of $\M_2$ associated to $x$ and $y$ respectively. It is not
difficult to see that, by putting $x_n=c_1\cdot\ldots\cdot c_n$ and
$y_n=d_1\cdot\ldots \cdot d_n$\,, one has:
\begin{align*}
  \bigvee_{n\geq1}x_n&=x\quad\text{in $\Mbar_1$}\,,
&\bigvee_{n\geq1}y_n&=y\quad\text{in $\Mbar_2$}\,.
\end{align*}

Then observe that the sequence $(x_n\cdot y_n)_{n\geq1}$ is non decreasing
in~$\M$. Indeed: 
\begin{align*}
  x_{n+1}\cdot y_{n+1}&\geq x_{n+1}\cdot y_{n}&&\text{by (\ref{item:5})
    and since $y_{n+1}\geq y_n$}\\
&=y_{n}\cdot x_{n+1}&&\text{by (\ref{item:4})}\\
&\geq
  y_n\cdot x_n&&\text{by (\ref{item:5})
    and since $x_{n+1}\geq x_n$}\\
&=x_n\cdot y_n&&\text{by (\ref{item:4})}\,.
\end{align*}

Since $\Mbar$ is complete w.r.t.\ the least upper bound of non
decreasing sequences (see~\cite[\S~2.1]{abbesmair14}), we define 
$f(x,y)\in\Mbar$ by:
\begin{gather*}
f(x,y)=\bigvee_{n\geq1}(x_n\cdot y_n) \quad\text{in $\Mbar$\,.}
\end{gather*}
It is then routine to see that $f$ thus defined is a bijection
$\Mbar_1\times\Mbar_2\to\Mbar$\,.

\medskip
Using the natural morphisms $\pi_1:\M\to\M_1$ and $\pi_2:\M\to\M_2$\,,
we have:
\begin{gather*}
  \forall x\in\M\qquad f^{-1}\bigl(\uup x\bigr)=\;\uup\bigl(\pi_1(x)
\bigr)\times\uup\bigl(\pi_2(x)\bigr)\,,\\
\forall (x_1,x_2)\in\M_1\times\M_2\qquad f(\uup x_1\times\uup
x_2)=\;\uup(x_1\cdot x_2)\,.
\end{gather*}
This shows that $f$ is a homeomorphism, hence \emph{a fortiori}
bi-measurable.

\medskip
Finally, let $\C$ denote the set of cliques of~$\M$, and let
$\C_1,\C_2$ denote the sets of cliques of $\M_1$ and $\M_2$
respectively. The isomorphism $f:\M_1\times\M_2\to\M$ induces by
restriction a bijection $\C_1\times\C_2\to \C$, from which follows the
product decomposition $\mu_\M=\mu_{\M_1}\times\mu_{\M_2}$\,. The proof
of Proposition~\ref{prop:1} is complete.

\par\bigskip\noindent\textbf{Proof of Proposition~\ref{thr:1}.}\quad
Assume for simplicity that $\M$ has two irreducible components, say
$\M_1$ and~$\M_2$\,. For $x\in\M$, let $x_1=\pi_1(x)\in\M_1$ and
$x_2=\pi_2(x)\in\M_2$ be the components of $x$ in $\M_1$ and
in~$\M_2$\,. Then we have:
\begin{align*}
  \nu_{p}(\up x)=p^{|x|}=p^{|x_1|}\times
  p^{|x_2|}=\nu'_{p}(\uup x_1)\times\nu''_{p}(\uup x_2)\,,
\end{align*}
where $\nu'_{p}$ is the sub-uniform measure of parameter $p$
on~$\Mbar_1$\,, and $\nu''_{p}$ is the sub-uniform measure of
parameter $p$ on~$\Mbar_2$\,. Using again that the collection of
elementary cylinders is a $\pi$-system, this is enough to conclude
that $\nu_{p}=\nu'_{p}\otimes\nu''_{p}$\,. The proof of
Proposition~\ref{thr:1} is complete.

\par\bigskip\noindent\textbf{Proof of Theorem~\ref{thr:8}.}\quad
Our proof extends the proofs of~\cite[Th.~2.4, Th.~2.5]{abbesmair14}
by taking into account the possible presence of the empty clique in
the Cartier-Foata decomposition of traces. It also makes a specific
use of the existence of the uniform measure obtained through the
construction of Theorem~\ref{thr:6}. Let $p$ be a real number such
that $p\leq p_0$\,.

For two cliques $c,c'\in\C$, let us write $c\parallel c'$ whenever
$c\times c'\subseteq I$, which is equivalent to saying both $c\cap
c'=\emptyset$ and $c\cdot c'\in\C$.

Let us denote by $C_1,\ldots,C_k$ the first $k$ cliques in the
Cartier-Foata decomposition of a random trace $\xi\in\Mbar$.  Let
$c_1\to\ldots\to c_k$ be a Cartier-Foata sequence of cliques,
$c_i\in\C$ for all $i\in\{1,\ldots,k\}$, and put
$x=c_1\cdot\ldots\cdot c_{k-1}$\,.  Then we have:
\begin{align}
\label{eq:6}
  \{\xi\in\Mbar\tq C_1=c_1,\ldots,C_k=c_k\}&=\uup(x\cdot
  c_k)\setminus\bigcup_{c\in\C\tq c>c_k}\uup(x\cdot c)
\end{align}

Let $(a_1,\ldots,a_r)$ be an enumeration of the elements $a$ of the
alphabet $A$ such that $a\notin c_k$ and $c_k\cup\{a\}\in\C$\,, or
equivalently with the above notation, those $a\in A$ such that
$a\parallel c$.  Then~(\ref{eq:6}) rewrites as:
\begin{gather*}
  \{\xi\in\Mbar\tq C_1=c_1,\ldots,C_k=c_k\}=\uup(x\cdot
  c_k)\setminus\bigcup_{j=1}^r\uup(x\cdot c_k\cdot a_j)
\end{gather*}

Passing to the probabilities on both sides yields:
\begin{gather}
\label{eq:7}
\nu_p(C_1=c_1,\ldots,C_k=c_k)=\nu_p\bigl(\uup(x\cdot
c_k)\bigr)-\underbrace{\nu_p\Bigl(\bigcup_{j=1}^r\uup(x\cdot c_k\cdot
  a_j)\Bigr)}_R\,,
\end{gather}
since the union on the right side member is included in $\uup(x\cdot
c_k)$. We evaluate the term $R$ in~(\ref{eq:7}) using Poincar\'e
inclusion-exclusion principle:
\begin{align*}
  R&=\sum_{j=1}^r(-1)^{j+1}\sum_{1\leq l_1<\cdots< l_j\leq
    r}\nu_p\bigl(\uup(x\cdot c_k\cdot a_{l_1})\cap\ldots\cap\uup(x\cdot c_k\cdot
  a_{l_j})\bigr)\\
  &=\sum_{j=1}^r(-1)^{j+1}\sum_{c'\in\Cstar\tq c_k\parallel
    c'\wedge |c'|=j}\nu_p\bigl(\uup (x\cdot c_k\cdot c')\bigr)\\
&=\sum_{c'\in\Cstar\tq c_k\parallel
  c'}(-1)^{|c'|+1}\nu_p(\bigl(\uup(x\cdot c_k\cdot c')\bigr)\\
&=\sum_{\delta\in\C\tq \delta>c_k}(-1)^{|\delta|-|c_k|+1}\nu_p\bigl(\uup
(x\cdot\delta)\bigr)\quad\text{with the change of variable $\delta=c_k\cdot c'$}
\end{align*}

Returning to~(\ref{eq:7}), we get:
\begin{align*}
  \nu_p(C_1=c_1,\ldots,C_k=c_k)&=p^{|x|+|c_k|}+\sum_{\delta\in\C\tq
    \delta>c_k}(-1)^{|\delta|-|c_k|}p^{|x|+|\delta|}\\
&=p^{|x|}\Bigl(\ \sum_{\delta\in\C\tq\delta\geq
  c_k}(-1)^{|\delta|-|c_k|}p^\delta\Bigr)\,.
\end{align*}
From this we deduce the following formula:
\begin{gather}
  \label{eq:8}
  \nu_p(C_1=c_1,\ldots,C_k=c_k)=p^{|x|}h(c_k)\,,\qquad\text{with
    $x=c_1\cdot\ldots\cdot c_{k-1}$\,,}
\end{gather}
if $c_1\to\ldots\to c_k$ holds, and where $h:\C\to\RR$ is defined
by~(\ref{eq:18}). In particular for $k=1$, we get:
\begin{gather*}
\forall c_1\in\C\qquad\nu_p(C_1=c_1)=h(c_1)\,,
\end{gather*}
which proves at once that $\bigl(h(c)\bigr)_{c\in\C}$ is a probability
vector, and that it is the distribution of the first clique~$C_1$
under~$\nu_p$\,.

Let us prove that $h$ is non zero on~$\Cstar$. Since $\M$ is
irreducible, the graph $(\Cstar,\to)$ of non empty cliques is strongly
connected (a well known result, see a proof in~\cite{krob03}). Let
$c\in\Cstar$, and let $c'\in\Cstar$ be maximal in $(\C,\leq)$. Let
$c_1,\ldots,c_n$ be $n\geq2$ non empty cliques such that
$c_1\to\ldots\to c_n$ holds, and $c_1=c$ and $c_n=c'$\,. Then,
by~(\ref{eq:8}), we have:
\begin{align*}
  \nu_p(C_1=c_1,\ldots,C_n=c_n)&=p^{|c_1|+\ldots+|c_{n-1}|}h(c_n)\\
&=p^{|c_1|+\ldots+|c_{n-1}|}p^{|c_n|}\,,
\end{align*}
since $h(c_n)=p^{|c_n|}$, by the maximality of~$c_n$\,. In particular:
\begin{gather*}
\nu_p(C_1=c_1)\geq\nu_p(C_1=c_1,\ldots,C_n=c_n)\neq0\,.
\end{gather*}
But we also have $\nu_p(C_1=c_1)=h(c_1)$, and thus $h(c_1)\neq0$,
which was to be shown. The value $h(1_\M)$ coincides with
$h(1_\M)=\mu_\M(p)$. Since $p_0$ is the root of smallest modulus
of~$\mu_\M$\,, and since $p\leq p_0$\,, it follows that $h(1_\M)=0$ if
and only if $p=p_0$\,.

We now come to the proof that $(C_k)_{k\geq1}$ is a Markov chain, and
to the computation of its transition matrix~$P$. If $p=p_0$, then this
is the result of~\cite[Th.~2.5]{abbesmair14}. The identity of the
transition matrices given in the present statement on the one hand,
and in~\cite[Th.~2.5]{abbesmair14} on the other hand, follows
from~\cite[Prop.~4.12]{abbesmair14}. Hence, assume that $p<p_0$\,.

If $c_1\to\ldots\to c_k$ holds, with all $c_j\neq0$, then the
expression~(\ref{eq:8}) combined with the fact that $h\neq0$
on~$\Cstar$\,, implies that $\nu_p(C_1=c_1,\ldots,C_{k-1}=c_{k-1})\neq
0$. Henceforth the following conditional probability is well defined:
\begin{align}
\label{eq:10}
  \nu_p(C_k=c_k|C_1=c_1,\ldots,C_{k-1}=c_{k-1})&=\frac{p^{|c_1|+\ldots+|c_{k-1}|}h(c_k)}{p^{|c_1|+\ldots+|c_{k-2}|}h(c_{k-1})}\\
\label{eq:11}
&=\frac{p^{|c_{k-1}|}h(c_k)}{h(c_{k-1})}
\end{align}

In case one of the $c_j$ is the empty clique, then the cliques
$c_{j+1},\ldots,c_k$ must also be empty since we assume that
$c_1\to\ldots\to c_k$ holds, and thus:
\begin{align}
\label{eq:9}
  \nu_p(C_k=1_\M|C_1=c_1,\ldots,C_{k-1}=c_{k-1})&=1
\quad\text{if one $c_j$ with $j<k$ is $1_\M$.}
\end{align}
Since $c_{k-1}=c_k=1_\M$, the right member of~\eqref{eq:11} evaluates
to~$1$ as well in this case. Hence~(\ref{eq:11}) is valid in all cases
if $c_1\to\ldots\to c_k$ holds. Since the right member
of~\eqref{eq:11} only depends on $c_{k-1},c_k$ on the one hand, and
since on the other hand it is clear that we have:
\begin{gather*}
  \nu_p(C_1=c_1,\ldots,C_k=c_k)=0\,,\quad\text{if $c_1\to\ldots\to
    c_k$ does \emph{not} hold,}
\end{gather*}
we conclude that $(C_k)_{k\geq1}$ is indeed a Markov chain with the
transition matrix described in the statement of the theorem.

\medskip For the proof of the converse part of the statement, we
proceed in four steps. Consider the two following claims:
\begin{enumerate}
\item\label{item:1} The vector\/ $\bigl(h(c)\bigr)_{c\in\C}$ is a probability
  vector.
\item\label{item:2} The matrix $P$ is stochastic.\setcounter{mylocalcounter}{\theenumi}
\end{enumerate}

Since we already know the existence of the measure~$\nu_p$ by
Theorem~\ref{thr:6}, the results already shown so far in the proof
make both points~\ref{item:1}--\ref{item:2} immediate. They follow
from the mere existence of the Markov chain $(C_k)_{k\geq1}$
previously defined under the measure~$\nu_p$ on~$\Mbar$, since $h$ is
the distribution of $C_1$ and $P$ is the transition matrix of the
chain.

For the next claim, we introduce new notations in order to avoid
confusion with the Markov chain $(C_k)_{k\geq1}$ previously defined.
\begin{enumerate}
  \setcounter{enumi}{\themylocalcounter}\item\label{item:3} Let $(C'_k)_{k\geq1}$ be a
  Markov chain on $\C$ with initial distribution $h$ and transition
  matrix~$P$, and let $Y'_k=C'_1\cdot\ldots\cdot C'_k$\,. Then the law
  of\/ $\bigvee_{k\geq1}Y'_k\in\Mbar$\; is~$\nu_p$\,.\setcounter{mylocalcounter}{\theenumi}
\end{enumerate}

Let $(\Omega,\GGG,\pr)$ be the sample space on which the Markov chain
$(C'_k)_{k\geq1}$ is defined, and put $\xi'=\bigvee_{k\geq1}Y'_k$ (see
the proof of Prop.~\ref{prop:1} above for the existence of the least
upper bound in~$\Mbar$)\,. Let also $\xi\in\Mbar$ be the canonical
random variable defined on $\Mbar$ with law~$\nu_p$\,. Then we have,
for every sequence $(c_1,\ldots,c_k)$ of cliques:
\begin{gather*}
  \pr(C'_1=c_1,\ldots,C'_k=c_k)=\nu_p(C_1=c_1,\ldots,C_k=c_k)\,,
\end{gather*}
since $(C_k)_{k\geq1}$ and $(C'_k)_{k\geq1}$ have same initial
distribution and same transition matrix. Therefore, for every $x\in\M$ and for
$k=\height(x)$, we have:
\begin{gather*}
  \pr(\xi'\geq x)=\pr(\cut k(\xi')\geq x)=\nu_p(\cut k(\xi)\geq
  x)=\nu_p(\xi\geq x)=\nu_p(\uup x)\,.
\end{gather*}




%
This proves that $\nu_p$ is indeed the law of~$\xi'$, and completes
the proof of Point~\ref{item:3}.

\smallskip Finally, it remains only to show the following point:

\begin{enumerate}
  \setcounter{enumi}{\themylocalcounter}\item The sequence
  $(Y'_k)_{k\geq1}$ converges toward $\nu_p$ in distribution.
\end{enumerate}

With the same notations as above, we have for every $x\in\M$ and for
every integer $k\geq\height(x)$:
\begin{align*}
  \pr(Y'_k\geq x)&=\pr\Bigl(\bigvee_{j\geq1}Y'_j\geq x\Bigr)=\pr(\xi'\geq
  x)=\nu_p(\uup x)\,,&\text{by Point~\ref{item:3}.}
\end{align*}

Hence for every $x\in\M$, the value of $\pr(Y'_k\geq x)$ is
eventually constant when $k\to\infty$, equal to $\nu_p(\uup x)$. This
implies the convergence of $(Y'_k)_{k\geq1}$ in distribution toward
the distribution~$\nu_p$\,. 

\medskip The proof of Theorem~\ref{thr:8} is complete.

\par\bigskip\noindent\textbf{Proof of Lemma~\ref{lem:2}.}\quad
We first show that $g$ is $B$-invariant on the right. For $p=p_0$\,,
it follows from~\cite[Prop.~4.12]{abbesmair14} that the formula
$h(c)=p_0^{|c|}g(c)$ holds for all cliques $c\in\Cstar$. Therefore,
for all $c\in\Cstar$, we have:
\begin{align*}
  (Bg)_c=\sum_{c'\in\C\tq c\to c'}p_0^{|c'|}g(c')=\sum_{c'\in\C\tq
    c\to c'}h(c')=g(c).
\end{align*}

We now prove that $B$ has spectral radius~$1$. Let $\|M\|$ denote the
spectral radius of a non-negative matrix, that is to say, the largest
modulus of its eigenvalues. For $p\leq p_0$, let $B_p$ be the matrix
of size $|\Cstar|$ and defined by
\begin{align*}
  (B_p)_{(c,c')}=
    \begin{cases}
      0,&\text{if $\neg(c\to c')$\,}\\
p^{|c'|},&\text{if $c\to c'$\,.}
    \end{cases}
\end{align*}
Hence: $B=B_{p_0}$\,. Let us show that:
\begin{gather}
\label{eq:26}
  \forall p<p_0\qquad\|B_p\|\leq1\,.
\end{gather}
Then by upper semi-continuity of the spectral radius, letting $p\to
p_0$\,, we will deduce $\|B\|\leq1$. And since we already proved that
$B$ has a right-invariant vector, we will obtain the equality
$\|B\|=1$.

To prove~(\ref{eq:26}), it is enough to show the following:
\begin{gather}
\label{eq:27}
\forall \lambda\in(0,1)\qquad I'\Bigl(\sum_{k\geq0}(\lambda B_p)^k\Bigr)
F<\infty\,,
\end{gather}
where $I$ and $F$ are the positive vectors of dimension $|\Cstar|$
defined as follows:
\begin{align*}
  \forall c\in\Cstar\qquad I_c&=1\,,&  \forall c\in\Cstar\qquad F_c&=\lambda p^{|c|}\,,
\end{align*}
and $I'$ is the transpose of~$I$. Indeed, (\ref{eq:27})~implies that
$\lambda^{-1}\Id-B_p$ is invertible for all $\lambda\in(0,1)$, and
thus, since $B_p$ is non-negative, that its largest eigenvalue cannot
be greater than~$1$.

Fix $\lambda\in(0,1)$. Then:
\begin{gather*}
\forall k\geq0\qquad  I'(\lambda B_p)^{k}F=\sum_{u\in\M\tq\height(u)=k}\lambda^{k}p^{|u|}\,,
\end{gather*}
and thus:
\begin{align}
\label{eq:28}
I'\Bigl( \sum_{k\geq0}(\lambda
B_p)^k\Bigr)F&=\sum_{k\geq0}\lambda^kR_p(k)\,,\\
\notag\text{with\quad}R_p(k)&=\sum_{u\in\M\tq\height(u)=k}p^{|u|}\,.
\end{align}
But, for $p<p_0$, we have:
\begin{align*}
  \sum_{k\geq0}R_p(k)&=\sum_{u\in\M}p^{|u|}=G(p)<\infty\,.
\end{align*}
Therefore $\lim_{k\to\infty}R_p(k)=0$. By~(\ref{eq:28}), it follows
that~(\ref{eq:27}) holds, which was to be shown.  The proof of
Lemma~\ref{lem:2} is complete.

\par\bigskip\noindent\textbf{Proof of Proposition~\ref{prop:2}.}\quad
Clearly, $C_{c,c'}=0=P_{c,c'}$ if $c\to c'$ does not hold. For
$c,c'\in\Cstar$ such that $c\to c'$ holds, we have:
\begin{align}
  \label{eq:5}
 C_{c,c'}=p_0^{|c'|}g(c')/g(c)=h(c')/g(c)\,,
\end{align}
by the formula $h(\cdot)=p_0^{|\cdot|}g(\cdot)$ recalled above in the proof of
Lemma~\ref{lem:2}. Still for $p=p_0$\,, we have, according to
formula~(\ref{eq:19}) of Theorem~\ref{thr:8}:
\begin{align}
  \label{eq:12}
P_{c,c'}&=h(c')p_0^{|c|}/h(c)=h(c')/g(c)\quad\text{since $h(c)=p_0^{|c|}g(c)$.}
\end{align}
Comparing~(\ref{eq:5}) and~(\ref{eq:12}), we obtain that $P=C$ on
$\Cstar\times\Cstar$\,. To get that $C$ is stochastic, it remains only
to show that $P_{c,1_\M}=0$ for all cliques $c\in\Cstar$. And indeed, by
formula~(\ref{eq:19}), we have:
\begin{gather*}
  \forall c\in\Cstar\quad P_{c,1_\M}=\frac{p_0^{|c|}h(1_\M)}{g(c)}
=\frac{p_0^{|c|}\mu_\M(p_0)}{g(c)}=0\,.
\end{gather*}
The proof of Proposition~\ref{prop:2} is complete.

\par\bigskip\noindent\textbf{Proof of Theorem~\ref{thr:9}.}\quad
Let $\nu$ denote the uniform measure on the boundary. We compute the
expectation of $\phibar(C_1\cdot\ldots\cdot C_k)$ under $\nu$ as
follows:
\begin{align*}
  \esp_\nu\phibar(C_1\cdot\ldots\cdot
  C_k)&=\sum_{x\in\M\tq\height(x)=k}\nu(C_1\cdot\ldots \cdot C_k=x)\biggl(\ \sum_{y\in\M\tq y\leq
    x\wedge|y|=k}\phi(y)\biggr)\\
&=\sum_{y\in\M\tq|y|=k}\phi(y)\biggl(\ 
\sum_{x\in\M\tq\height(x)=k\wedge x\geq y}\nu(C_1\cdot\ldots\cdot
C_k=x)\biggr)\\
&=\sum_{y\in\M\tq|y|=k}\phi(y)\nu(\up y)\\
&=p_0^k\cdot\bigl(\#\M_k\bigr)\cdot\esp_{\nu_{\M_k}}\phi\qquad\text{since
  $\nu(\up y)=p_0^{|y|}=p_0^k$}
\end{align*}

This completes the proof of Theorem~\ref{thr:9}.

\end{document}